\documentstyle[prl,aps,epsfig]{revtex}  
\bibstyle{unsrt}

\begin{document}   
\twocolumn[\hsize\textwidth\columnwidth\hsize\csname
@twocolumnfalse\endcsname
\draft

\title{Phase Control of Nonadiabaticity-induced Quantum
Chaos in An Optical Lattice}
\author{Jiangbin Gong and Paul Brumer}
\address{ Chemical Physics Theory Group,
University of Toronto, Toronto, Canada  M5S 3H6}
\date{\today}
\maketitle

\begin{abstract}              
The qualitative nature (i.e. integrable vs. chaotic) of the translational
dynamics of a three-level atom in an optical lattice
is shown to be controllable by varying the relative laser phase of two
standing wave lasers. Control is explained in terms of
the nonadiabatic transition between optical potentials and the corresponding
regular to chaotic transition in mixed classical-quantum dynamics.
The results are of interest to both areas of coherent control and quantum 
chaos.
\end{abstract}

\vskip1pc]

Recent years have witnessed an increasing interest in
the coherent control of atomic and molecular processes \cite{brumer00,rice01}.
One central aspect of coherent control is {\em phase control}, in which 
different optical phases are introduced into coherent 
laser-matter interactions in order to manipulate
quantum interference effects and thus to achieve target
objectives.  It has been shown that phase control approaches are
widely applicable \cite{brumer00}, even to some systems displaying
quantum chaotic dynamics\cite{gong01}.

As we show in this letter, optical lattices, of great recent 
interest\cite{raizenetc}, provide an important system in which to explore
aspects of quantum chaos and coherent control. In particular, we describe
the all-optical phase control of the translational motion of atoms in a system
that allows chaotic vs. integrable motion depending on the phase shift between  
two standing-wave laser fields. 
The control mechanism is shown to originate in
the nonadiabatic coupling between different optical potentials as well as
in the regular to chaotic transition
in a mixed classical-quantum description of the model system.
The results are of broad interest to  both coherent control and
quantum chaos.

Consider a $\Lambda$-type 3-level atom moving along
two co-propagating standing-wave laser beams, with two lower degenerate levels
$|1\rangle$ and $|3\rangle$, and one upper level $|2\rangle$.
Two laser fields, with different polarizations $\sigma_{+}$ and $\sigma_{-}$, 
couple $|1\rangle$ with $|2\rangle$ and $|2\rangle$ with $|3\rangle$, 
respectively. A closed 3-level $\Lambda$ configuration of this kind may be 
realized, for example, in $^{4}$He using the 
$2^{3}S_{1}\rightarrow 2^{3}P_{1}$ transition \cite{aspect}. 
The laser fields are of the same frequency,
with large detuning $\Delta$ from $|2\rangle$.
We use $x$, $p$, $M$, $\Omega_{1}$ ($\Omega_{2}$),
$k_{1}$ ($=k_{2}$) to
represent the position, momentum, atomic mass, the 
two Rabi frequencies (assumed real) and the two wavevectors,
respectively. The relative phase of the two
standing-waves is denoted by $\phi$. 
For generality we employ a set of natural units by scaling 
all the parameters,  i.e., $x^{0}=\lambda$ for $x$, $p^{0}=
\hbar /\lambda$ for $p$, $t^{0}=M\lambda^{2}/\hbar$
for the time variable $t$, $\Omega^{0}=2/t^{0}$ for
$\Delta$, $\Omega_{1}$ and $\Omega_{2}$. In terms of
these units, the dynamical equations do not explicitly contain the
atomic mass, the effective wavevector is given by $k=2\pi$,
and $[x,p]=i$ in the full quantum dynamics.
In the rotating wave approximation and in the interaction picture,
the Hamiltonian describing the translational motion along the laser beams
is $H=p^{2}/{2}+(V_{ij})$, with the potential matrix $(V_{ij})\ (i,j=1,2,3)$ 
given by
\begin{equation}
(V_{ij}) = \left ( 
\begin{array}{ccc}
 2\Delta & \Omega_{1}\sin(kx) & 0 \\
 \Omega_{1}\sin(kx) & 0 & \Omega_{2}\sin(kx+\phi) \\
 0 & \Omega_{2}\sin(kx+\phi) & 2\Delta \\
\end{array} \right ).
\label{potentmatrix}
\end{equation}

\vspace{-0.4cm}
\begin{figure}[ht]
\begin{center}
\epsfig{file=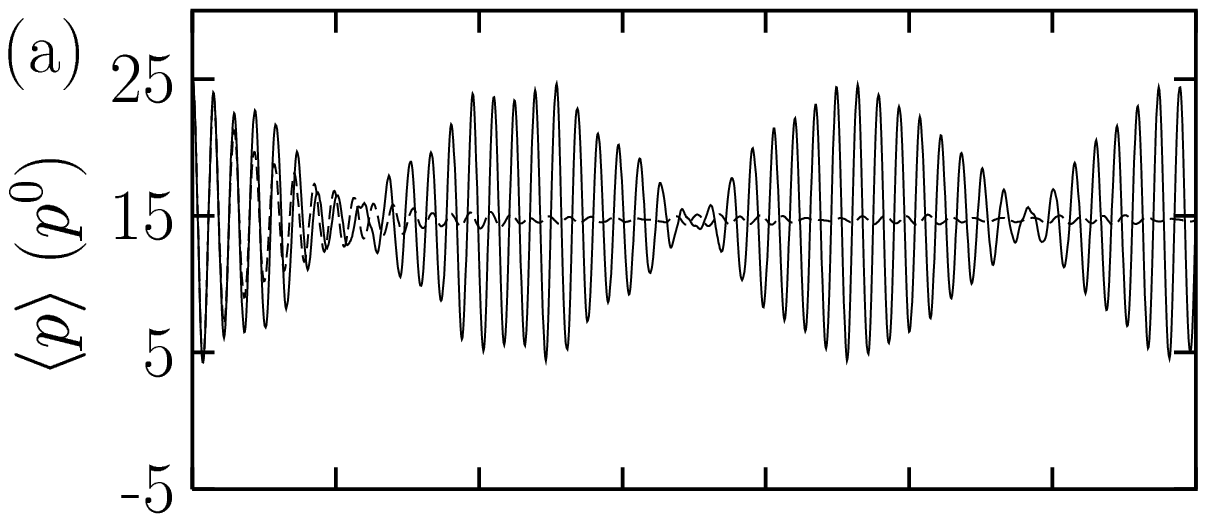,width=7cm}
\vspace{-1.3cm}

\epsfig{file=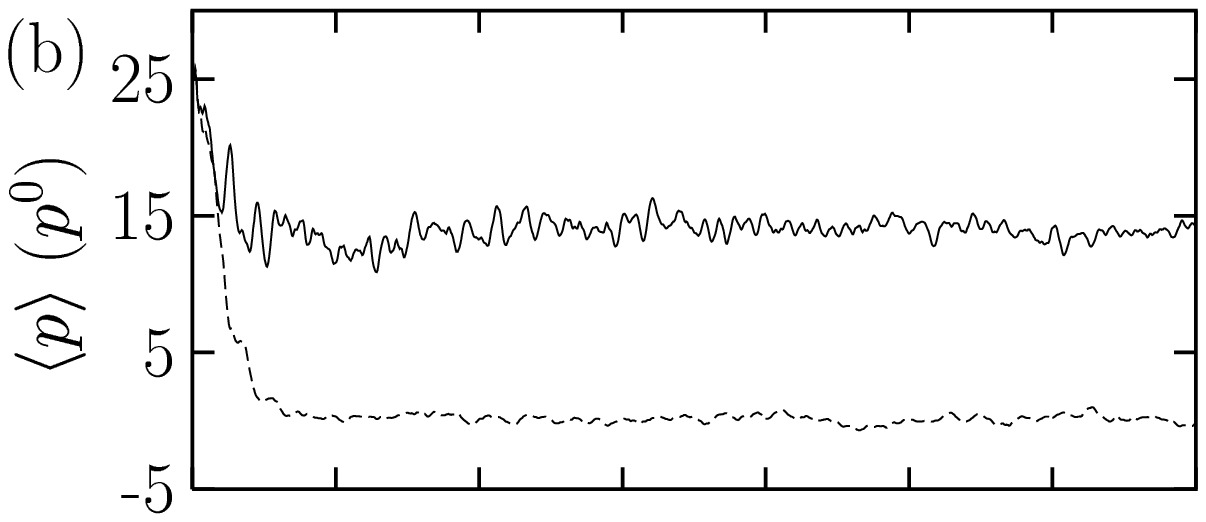,width=7cm}
\vspace{-1.3cm}

\epsfig{file=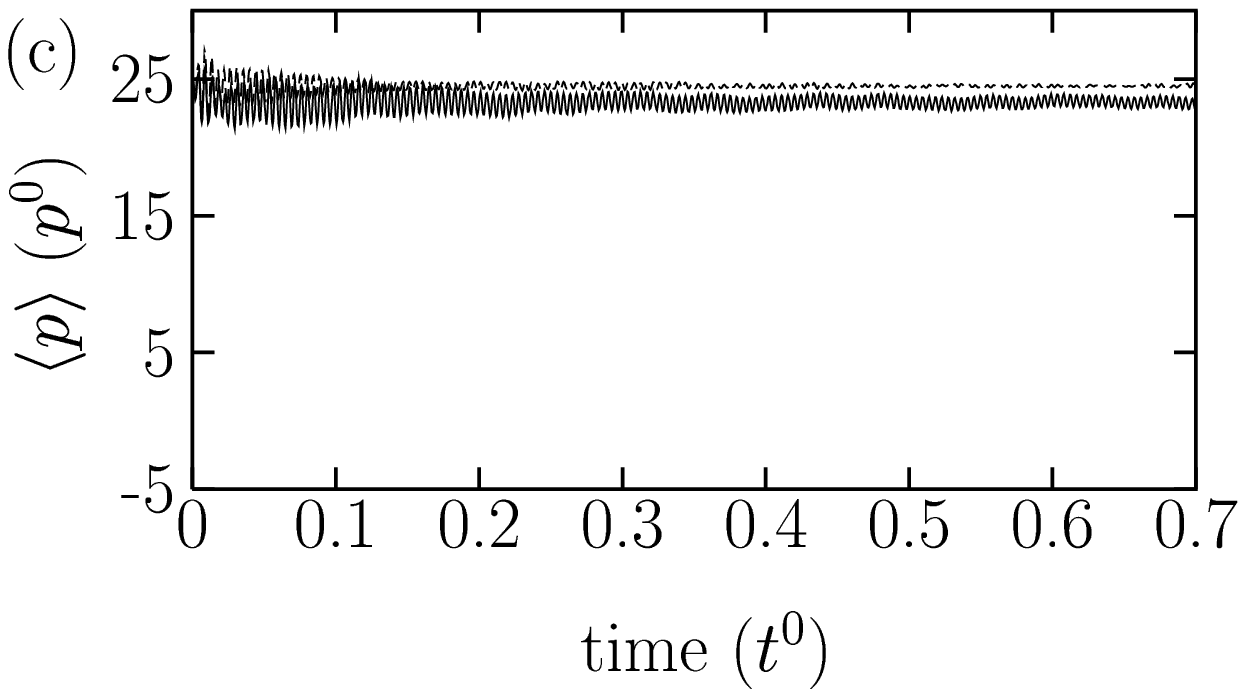,width=7cm}
\end{center}
\vspace{-0.2cm}
\caption{
Time dependence of $\langle p \rangle$
obtained from  quantum wavepacket dynamics calculations (solid lines) and
the ensemble statistics in a mixed classical-quantum description (dashed lines).
The initial internal state is $|1\rangle$,
and the $\langle x\rangle$ and $\langle p \rangle$
of the initial Gaussian ensemble are
0.0 and 25.0, respectively. The
initial variances in position
and momentum are chosen to be $1.0/10\sqrt{2}$ and $10.0/\sqrt{2}$,
respectively. The relative phase $\phi$
equals
$0.0$, $0.25\pi$, and $0.5\pi$ in (a), (b) and (c), respectively.
}
\label{fig-exp}
\end{figure}

To demonstrate phase control of the dynamics, consider numerical results
for the specific case where (1) $\Omega_{1}=6.0\times 10^{3}$,  $\Omega_{2}=7.0\times
10^{3}$, $\Delta=1.5\times 10^{4}$; (2) the atom is initially
in internal state $|1\rangle$; and (3) the average momentum, average position,
momentum variance, and position variance of the initial Gaussian wavepacket
are given by $\langle p\rangle =25.0$, $\langle x\rangle=0.0$,
$\delta p=10/\sqrt{2}$ and $\delta x=1.0/10\sqrt{2}$, respectively.  In 
$^{4}$He, this corresponds to $t^{0} \sim$ 77 $\mu$sec, 
the detuning $\sim$ $2\pi \cdot 62$ MHz (about 38 times the
linewidth of $|2\rangle$), and  the initial kinetic temperature 
$\sim$ 29 $\mu$K. Results for this case represent typical
observations for a wide range of system parameters and initial conditions
that we have examined\cite{para-note}.

The solid curves in Fig. \ref{fig-exp}, which contain the essential result of
this letter, show the time dependence of the quantum momentum expectation
value $\langle p \rangle $ for time evolving wavepackets for various
$\phi$. The $\phi=0$ case [Fig. \ref{fig-exp}a] displays a perfectly regular 
recurrence pattern. By contrast, the  $\phi=0.25\pi$ case [Fig. \ref{fig-exp}b]
is characteristic of a relaxation process, with significantly 
irregular oscillations of small amplitude. In addition,  the associated power 
spectrum (not shown) is quite noisy, characteristic of chaos\cite{dumont}.   
Further tuning $\phi$ leads to totally different dynamics: 
in the $\phi=0.5\pi$ case [Fig. \ref{fig-exp}c], $\langle p\rangle$
lies very close to its initial value and
undergoes regular oscillations,  but with a characteristic 
frequency that is much higher than that in the $\phi=0$
case (The dashed curves in Fig. \ref{fig-exp} are discussed later below).
Clearly, the atom's translational motion undergoes significant qualitative
changes with controlled changes in $\phi$.

\begin{figure}[ht]
\begin{center}
\epsfig{file=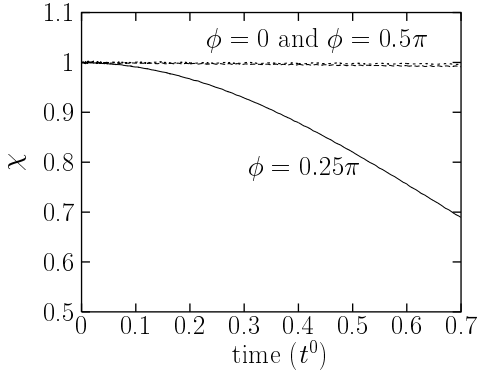,width=7cm}
\vspace{-0.4cm}
\caption{
The sensitivity of the quantum dynamics to slight changes of the relative phase
parameter $\phi$. $\chi$ is the absolute value of
the overlap between the two time evolving wavefunctions emanating from
the same initial state as in Fig. \ref{fig-exp}, with the relative phase of the two
standing-wave
laser fields
given by $\phi$ and $\phi+\pi/400$.
}
\label{fig-overlap}
\end{center}
\end{figure}

To demonstrate that this is indeed an integrable to chaotic transition, we
consider the sensitivity of the dynamics to slight changes of $\phi$.
Figure \ref{fig-overlap} shows the time dependence of the absolute value of 
the overlap $\chi$ of two time evolving wavefunctions
emanating from the same initial Gaussian wavepacket, with the relative
phase of the two laser fields given by  $\phi$
and $\phi+\pi/400$, respectively. For $\phi=0$ or $\phi=0.5\pi$,  $\chi$ 
remains near unity throughout, indicating that
the dynamics is insensitive to tiny changes of $\phi$.  By contrast, for
$\phi=0.25\pi$, $\chi$ is already less than 0.70 at $t=0.7$.  This
interesting  {\em hypersensitivity} \cite{caves} to perturbations in the 
relative phase parameter $\phi$ 
further suggests that the $\phi=0.25\pi$ case is indicative of quantum
chaos.

\begin{figure}[ht]
\hskip 1.0 cm
\begin{center}
\epsfig{file=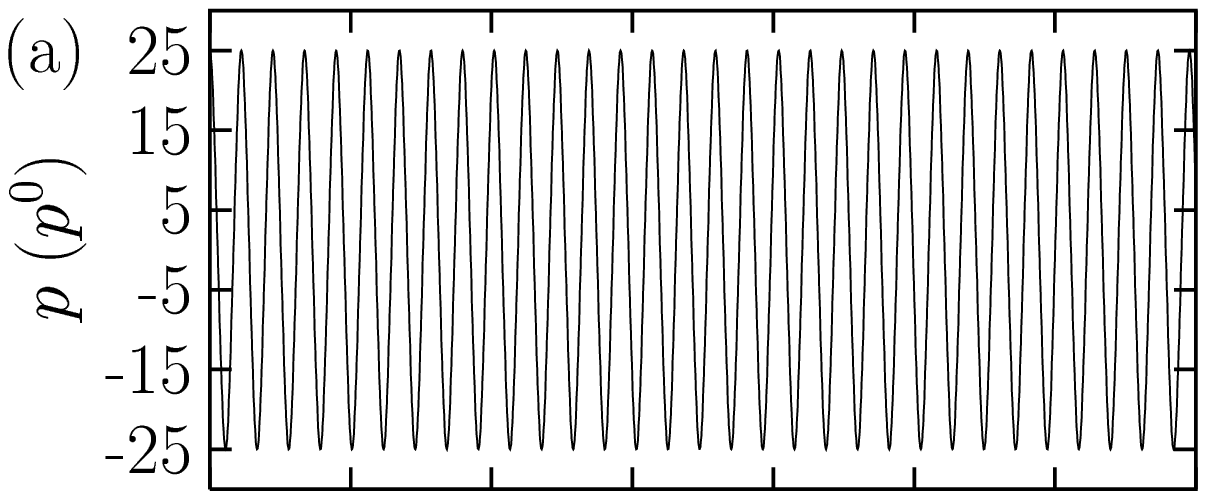,width=7cm}
\vspace{-1.3cm}

\epsfig{file=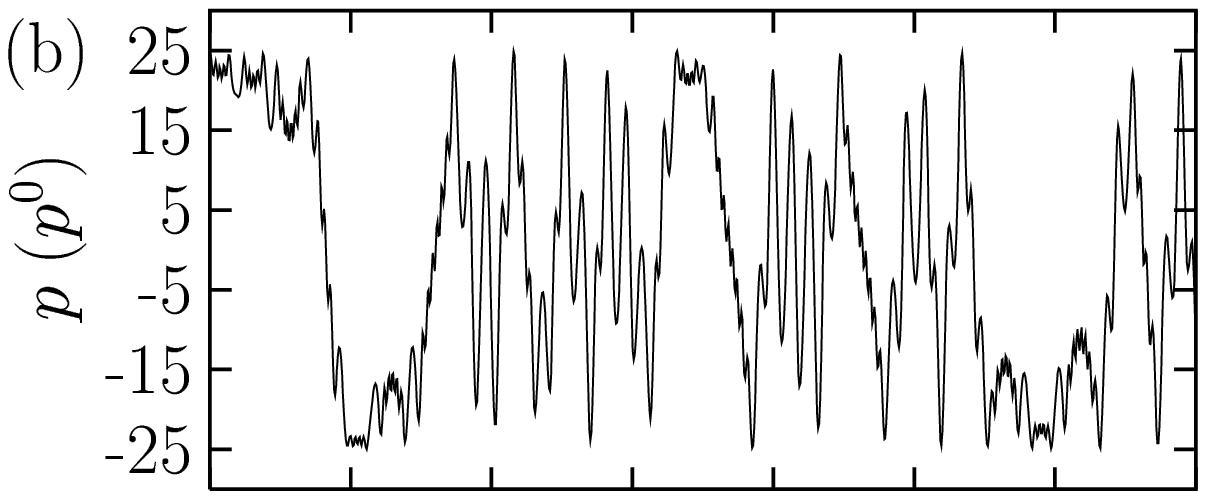,width=7cm}
\vspace{-1.3cm}

\epsfig{file=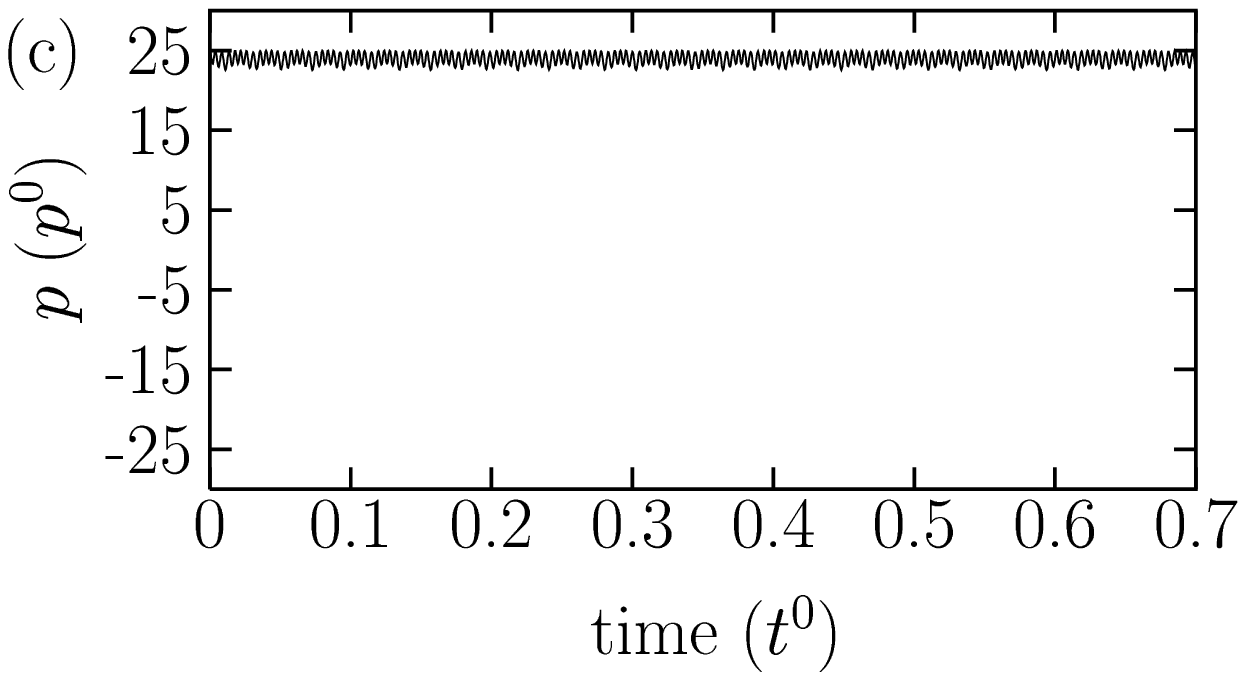,width=7cm}
\end{center}
\caption{
Time dependence of momentum for classical trajectories
obtained by solving Eqs. (\ref{mixed-eq1}), (\ref{mixed-eq2}), and
(\ref{mixed-eq3}), with
the initial position
$x=0$ and the initial momentum $p=25.0$.  The initial internal state is given by $|1\rangle$.
$\phi$ equals  $0.0$, $0.25\pi$, and $0.5\pi$ in (a), (b) and (c), respectively.
}
\label{fig-traj}
\end{figure}

To further substantiate that this is a regular to chaotic transition we
consider a mixed classical-quantum description of the dynamics, i.e., 
where the center-of-mass motion evolves
classically on an average potential and where
the internal motion is treated quantum mechanically \cite{tully,blumel}.
From this perspective, the state of motion is described by a phase space point
$(x, p)$ and an internal wavefunction  
$|\psi\rangle=(C_{1}, C_{2}, C_{3})$, where $C_{1},\ C_{2}$, and $C_{3}$ are the projections
of the wavefunction onto the three internal levels. 
The dynamical equations are given by 
\begin{equation}
\frac{dx}{dt}   =   p ,
\label{mixed-eq1}
\end{equation}
\begin{eqnarray}
 \frac{dp}{dt} & = &-\langle \psi|\frac{dV}{dx}|\psi\rangle \nonumber \\
& = &
-\Omega_{1}k(C_{2}C_{1}^{*}+C_{2}^{*}C_{1})\cos(kx) \nonumber \\
&&-\Omega_{2}k(C_{2}C_{3}^{*}+C_{2}^{*}C_{3})\cos(kx+\phi),
\label{mixed-eq2}
\end{eqnarray}
and
\begin{equation}
i\frac{dC_{k}}{dt}=\sum_{j=1}^{3}V_{kj}C_{j}, \ k=1,\ 2,\ 3.
 \label{mixed-eq3}
 \end{equation}

One can readily solve these equations
 numerically to obtain $p(t)$  for the classical
translational motion.  The results, for the same system parameters and 
the same initial internal state as in the quantum
calculations, but for a single trajectory initially at $(x,p)=(0.0, 25.0)$, 
are shown in  Fig. \ref{fig-traj}. In particular, for $\phi=0$ 
[Fig. \ref{fig-traj}a] the oscillation of momentum is perfectly regular, 
in agreement with the regular recurrence pattern in Fig. \ref{fig-exp}a.
By contrast, for $\phi=0.25\pi$ [Fig. \ref{fig-traj}b], the
trajectory is highly irregular, with random alternations between 
fast small-amplitude and  slow large-amplitude oscillations. In this case
two initially nearby trajectories shows exponential divergence in phase space 
with an associated Lyapunov exponent of $\sim$ $50/t^{0}$.  
This supports the view that the irregular dynamics in Fig. \ref{fig-exp}b,
and the sensitivity of the dynamics to slight changes of $\phi$
shown in Fig. \ref{fig-overlap}, are indeed due to optical-phase-assisted 
quantum chaos. Further, in the $\phi=0.5\pi$ case [Fig. \ref{fig-traj}c], the 
regular classical motion is restored, with a characteristic frequency 
identical to that in Fig. \ref{fig-exp}c. Hence optical-phase control is
evident in this classical-quantum treatment as well.

An ensemble statistics in the classical treatment of 
translational motion provides further support.
The dashed lines in Fig. \ref{fig-exp} display the time dependence of
the average momentum $\langle p \rangle$
for an ensemble of trajectories initially centered at $x=0$ and
$p=25.0$, with the same initial variances as in the quantum calculations.
Each individual trajectory in the ensemble
is obtained by solving Eqs. (\ref{mixed-eq1}) -- (\ref{mixed-eq3}).
The quantum-classical correspondence for the regular dynamics at times $t<0.15$
in Figs. \ref{fig-exp}a and \ref{fig-exp}c is impressive.
On the other hand, as seen in  Fig. \ref{fig-exp}b, $\langle p\rangle$
for the classical ensemble quickly relaxes to zero, whereas $\langle p\rangle$ for
the quantum ensemble remains far away from zero. This quantum-classical difference
constitutes an excellent example of quantum
suppression of classical chaos in an unbounded Hamiltonian system.

Insight into the origin of phase control
can be obtained by considering the dynamics in an
adiabatic representation. To do so we introduce
an orthogonal transformation $(O_{ij})\ (i,j=1,2,3)$ 
to diagonalize the potential matrix $(V_{ij})$. $(O_{ij})$ is given by
\begin{large}
\begin{equation}
(O_{ij})=\left (
\begin{array}{ccc}
\frac{\Omega_{1}\sin{kx}}{\sqrt{2(\eta^{2}-\Delta\eta)}} &
\frac{\Omega_{2}\sin(kx+\phi)}{\xi} & \frac{\Omega_{1}\sin(kx)}{
\sqrt{2(\eta^{2}+\Delta\eta)}} \\
\frac{\eta-\Delta}{\sqrt{2(\eta^{2}-\Delta\eta)}} & 0 & 
\frac{-\eta-\Delta}{\sqrt{2(\eta^{2}+\Delta\eta)}} \\
\frac{\Omega_{2}\sin(kx+\phi)}{\sqrt{2(\eta^{2}-\Delta\eta)}}
& \frac{-\Omega_{1}\sin(kx)}{\xi} & \frac{\Omega_{2}\sin(kx+\phi)}
{\sqrt{2(\eta^{2}+\Delta\eta)}}\\
\end{array}
\right ),
\label{omatrix}
\end{equation}
\end{large}
where
$\xi(x,\phi)\equiv$$\sqrt{
\Omega_{1}^{2}\sin^{2}(kx)+\Omega_{2}^{2}\sin^{2}(kx+\phi)}$
and $\eta(x,\phi)$$\equiv \sqrt{\xi^{2}(x,\phi)+\Delta^{2}}$.
Corresponding to the three eigenvectors $(O_{1j}, O_{2j}, O_{3j})$ ($j=1,2,3$),
are three eigen-potentials $V_{i}(x,\phi)$ given by
$V_{1}(x,\phi)=\Delta + \eta (x, \phi)$,  
the constant potential $V_{2}=2\Delta$,   
and $ V_{3}(x,\phi)=\Delta - \eta (x, \phi)$. 
For the special case of $\phi=0$, $V_{1}$ and $V_{2}$ are degenerate at $kx=n\pi$, and
the eigenvector $(O_{12}, O_{22}, O_{32})$ 
is $x$-independent; for the general cases of
$\phi\neq 0$,
$V_{1}(x,\phi)> V_{2}>V_{3}(x,\phi)$, i.e., the three potential curves
do not cross one another.   Of particular interest is
the constant potential $V_{2}$,
associated with the eigenvector $(O_{12}, O_{22}, O_{32})$.
Since
\begin{equation}
\Omega_{1}\sin(kx)O_{12}+\Omega_{2}\sin(kx+\phi)O_{32}=0, 
\end{equation}
$V_{2}$ results from 
the complete quantum
destructive interference between the two standing-wave laser fields. 
As such,  
$V_{2}$ is an extension of
the ``dark optical
lattice''
in the presence of
two counter-propagating plane-wave laser beams \cite{aspect,dum,dutta}.

Consider now transforming
Eqs. (\ref{mixed-eq2}) and (\ref{mixed-eq3}) to the eigen-potential (adiabatic) 
representation. Specifically,
consider the dynamics in terms of $\tilde{C}_{i},\ i=1,2,3$,
where 
$\tilde{C}_{i}=\sum^{3}_{k=1}O_{ki}{C}_{k}$.
Using  
$\sum^{3}_{k=1}O_{ki}O_{kj}=\delta_{ij}$ and 
$\sum^{3}_{k=1}O_{ki}dO_{kj}/dx=
-\sum^{3}_{k=1}O_{kj}dO_{ki}/dx$,
Eqs. (\ref{mixed-eq2}) and (\ref{mixed-eq3})  
can be transformed to 
\begin{equation}
\frac{dp}{dt}=-|\tilde{C}_{1}|^{2}\frac{dV_{1}(x,\phi)}{dx}
-|\tilde{C}_{3}|^{2}\frac{dV_{3}(x,\phi)}{dx}, 
\label{newmixed-eq2}
\end{equation}
and
\begin{equation}
i \left (
\begin{array}{c} \frac{d\tilde{C}_{1}}{dt}\\
\frac{d\tilde{C}_{2}}{dt}\\
\frac{d\tilde{C}_{3}}{dt}\\ \end{array}
\right )=\left (
\begin{array}{ccc} V_{1}(x,\phi) & it_{12}
& 
-it_{13}\\
-it_{12} & V_{2}&
-it_{23}\\
it_{13} &
it_{23}
& V_{3}(x,\phi)\\
\end{array}
\right )
\left (
\begin{array}{c}
\tilde{C}_{1}\\
\tilde{C}_{2}\\
\tilde{C}_{3}\\
\end{array}
\right ),
\label{newmixed-eq3}
\end{equation}
where the diagonal terms
are the three adiabatic potentials given above, and the potential coupling
terms are given by
$t_{12}(x,\phi)=
pk\Omega_{1}\Omega_{2}
\sin(\phi)/[2\xi^{2}(\eta^{2}-\Delta
\eta)]^{1/2}$, $t_{13}(x,\phi)=
pk\Delta[\Omega_{1}^{2}\sin(2kx)+\Omega_{2}^{2}\sin(2kx+2\phi)]/4\eta^{2}\xi$,
and $t_{23}(x,\phi)=
pk\Omega_{1}\Omega_{2}\sin(\phi)/[2\xi^{2}(\eta^{2}+\Delta
\eta)]^{1/2}$ \cite{note}.  
Note that $t_{ij}\ (i\neq j=1,2,3)$ is proportional to the momentum $p$. Thus, 
the coupling between optical potentials is due to the nonadiabatic
effects associated with translational motion.
Further,
for the computational example discussed above (and many other cases in which $\Delta>0$)
one has $\min[V_{1}(x,\phi)-V_{2}]<<\min[V_{2}-V_{3}(x,\phi)]$ and
$t_{12}(x,\phi)>>
t_{23}(x,\phi)$, suggesting that
$V_{3}(x,\phi)$ is effectively
decoupled from $V_{1}(x,\phi)$ and $V_{2}$. Note also that
at the initial location  $x=0$, the initial internal state $|1\rangle$ is a superposition state of 
the two eigenvectors associated with $V_{1}$ and $V_{2}$ for
$\phi=0$, and reduces to the eigenvector associated with
$V_{2}$  for $\phi\neq 0$.

The key role of the relative laser phase $\phi$ becomes clear as one
compares the magnitude of the nonadiabatic
coupling term $t_{12}(x,\phi)$ with that of  $(V_{1}-V_{2})$.
For case (a),  $\phi=0$ and $t_{12}(x,\phi=0)=0$, so the dynamics is adiabatic.
Thus,
in the case of Fig. \ref{fig-exp}a, the quantum ensemble divides into two sub-ensembles:
one is trapped in one well of $V_{1}$ around $x=0$ and
undergoes periodic oscillations, and the other experiences
the trivial motion on the constant potential $V_{2}$.   
For case (b), $\phi=0.25\pi$. Here the smallest gap between $V_{1}$ and $V_{2}$ 
is given by $g(\phi)=\sqrt{A^{2}(\phi)+\Delta^{2}}-\Delta$, where
$A^{2}(\phi)=[\Omega_{1}^{2}+\Omega_{2}^{2}-\sqrt{\Omega_{1}^{4}+
\Omega_{2}^{4}+2\Omega_{1}^{2}\Omega_{2}^{2}\cos(2\phi)}]/2$.  The corresponding
ratio of the potential coupling term $t_{12}$ 
to $g(\phi)$
is given by 
\begin{equation}
T(\phi)=\frac{pk\Omega_{1}\Omega_{2}\sin(\phi)}{\sqrt{2}[A^{2}(\phi)
+\Delta^{2}]^{1/4}A(\phi)
g^{3/2}(\phi)}.
\label{teq}
\end{equation}
Since $T(\phi=0.25\pi)\approx 1.0$, the magnitude of the nonadiabatic coupling 
is comparable to that of $(V_{1}-V_{2})$, resulting in strong
nonadiabatic effects.  Thus,  the chaotic motion seen in Fig. 
\ref{fig-traj}b is  induced by the significant nonadiabatic coupling between the two simple
one-dimensional 
potentials $V_{1}$ and $V_{2}$.
Finally, for case (c), $\phi=0.5\pi$.  Here $T(\phi=0.5\pi)=\min[T(\phi)]\approx 0.16$, i.e.,
the nonadiabatic coupling is appreciably weaker than in the case of $\phi=0.25\pi$.
As such, the translational motion, 
initially launched on the adiabatic potential $V_{2}$,
would essentially remain on $V_{2}$, with small perturbations  from  
the insignificant Rabi population oscillation between  $V_{1}$ and $V_{2}$. 
To further confirm
this picture, one finds that 
the characteristic frequency of the regular dynamics
in Fig. \ref{fig-exp}c and Fig. \ref{fig-traj}c is $\sim$ 1425, a value consistent with
the Rabi frequencies given by $\sqrt{(V_{1}-V_{2})^{2}+
4t_{12}^{2}}$ [see Eq. (\ref{newmixed-eq3})]. 

A number of additional remarks are in order. First,
the two-standing-wave configuration is essential in this system. That is, if
either or both of the two standing-wave fields are replaced by
a traveling-wave,  the nonadiabatic coupling or the spacing between
$V_{1}$ and $V_{2}$ is no longer a sensitive function of $\phi$, and there is
no significant phase control.  On the other hand, a  three-level atom in 
two standing-wave laser fields of different but commensurate frequencies 
also shows dynamics which is controllable by changing $\phi$ \cite{gong02}.
Second, in contrast to some recent 
studies on optical-magneto lattice \cite{jessen,ghose},  
the coupling between different optical potentials discussed above is not
due to additional magnetic fields, but directly due to nonadiabaticity.
Further, unlike the work in Ref. \cite{ghose}, here we have observed clear signatures of
quantum chaos in the quantum dynamics.
Thus, this model is the first all-optics realization
of nonadiabaticity-induced quantum chaos, a phenomenon first discovered in molecular systems \cite{heller}. 
Third, in this work we have neglected decoherence effects (e.g., 
due to the spontaneous emission from the excited state $|2\rangle $).  
It would be interesting to explore how decoherence affects phase control and
quantum-classical correspondence in this system.

In conclusion, we have demonstrated
optical phase control of nonadiabaticity-induced quantum chaos
in a $\Lambda$-type 3-level system in a two-standing-wave optical lattice.
The results shown in Fig. \ref{fig-exp} are but samples of the observable,
controllable, behavior. Further,
the functional dependence on $\phi$ has been exposed analytically, which can 
serve to guide
experimental studies of the $\phi$-dependent regular to chaotic transition.
Recent experimental progress in atom optics and quantum chaos
\cite{raizenetc,aspect,dutta,jessen,raizen} suggests that the results should be 
experimentally achievable with existent technology.
This work was supported by the U.S. Office of
Naval Research and the Natural Sciences and Engineering
Research Council of Canada.  J.G. is a Henry Croft Postdoctoral Fellow in
Theoretical Chemical Physics.


\begin{references}
\bibitem{para-note}The system parameters and 
 the initial conditions should be chosen to ensure that the population on 
 level $|2\rangle$ is small ($<$ 2\% in the present case) so that spontaneous
 emission effects will be insignificant.  It is also desirable to ensure
 negligible tunneling effects between different wells of $V_{1}$ 
 so as to achieve good quantum-classical correspondence.
\bibitem{note}
 The mixed
 classical-quantum treatment automatically neglects
 the optical Gauge potential \cite{dum,dutta}, which is not of much interest
 as far as the nonadiabatic coupling is concerned.
\end{references}
  \end{document}